\documentclass[11pt]{article}
\usepackage[a4paper]{geometry}
\usepackage[T1]{fontenc}
\usepackage[utf8]{inputenc}                 
\usepackage[USenglish]{babel}
\usepackage{emptypage}
\usepackage{indentfirst}
\usepackage{booktabs}  
\usepackage{tabularx} 
\usepackage{listings} 
\usepackage[font=small]{quoting}  
\usepackage{amsmath,amssymb,amsthm}                                   
\usepackage{chngpage,calc}   
\usepackage[dvipsnames]{xcolor}    
\usepackage{lipsum}
\usepackage{eurosym} 
\usepackage{xr-hyper}
\usepackage{hyperref} 
\usepackage{indentfirst}
\usepackage{authblk}
\usepackage{textcomp}
\usepackage{verbatim}
\usepackage{varwidth}
\usepackage{bbm}
\usepackage{url}

\usepackage{natbib}

\usepackage[noend]{algpseudocode}
\usepackage{dsfont}
\usepackage{float}
\usepackage[caption = false]{subfig}
\usepackage[final]{graphicx}
\usepackage{tasks}
\usepackage{courier}
\usepackage{gensymb}
\usepackage{multirow}
\usepackage[shortlabels]{enumitem}
\usepackage{gensymb}
\usepackage{bm}

\usepackage{marginnote}
\reversemarginpar

\usepackage{graphicx}

\makeatletter
\def\BState{\State\hskip-\ALG@thistlm}
\makeatother

\usepackage{titlesec}
\usepackage[dvipsnames]{xcolor}
\usepackage{epstopdf}
\usepackage{wrapfig}
\usepackage{verbatim}
\usepackage{pbox}
\usepackage{fancyhdr}
\usepackage{float}
\usepackage{adjustbox}
\usepackage{tikz}

\makeatletter
\newcommand*{\centerfloat}{%
	\parindent \z@
	\leftskip \z@ \@plus 1fil \@minus \textwidth
	\rightskip\leftskip
	\parfillskip \z@skip}
\makeatother
\renewcommand{\arraystretch}{1}
\usepackage{acro}
\usepackage{setspace}

\acsetup{first-style=short}
\usepackage{ragged2e}

\usepackage{minipage-marginpar}
\usepackage{placeins}
\usepackage{float}

\usepackage{listings}
\usepackage{afterpage}

\usepackage{algorithm2e}

\usepackage[flushleft]{threeparttable}

\usepackage{graphicx}

\usepackage{mathrsfs,upref}

\theoremstyle{definition}

\theoremstyle{remark}

\usepackage{enumitem}
\newlist{steps}{enumerate}{1}
\setlist[steps, 1]{label = \textbf{Step \arabic*}:}

\makeatletter
\newcommand*{\rom}[1]{\expandafter\@slowromancap\romannumeral #1@}
\makeatother
\numberwithin{equation}{section}

\newlength\myindent
\usepackage{verbatimbox}

\title{Quantile regression for longitudinal functional
data with application to feed intake of lactating sows}

 \author[1,*]{Maria Laura Battagliola}
 \author[2]{Helle S\o rensen}
 \author[2]{Anders Tolver}
 \author[3]{Ana-Maria Staicu}
\affil[1]{School of Basic Sciences, École Polytechnique Fédérale de Lausanne\\ \smallskip}
\affil[2]{Department of Mathematical Sciences, University of Copenhagen\\ \smallskip}
\affil[3]{Department of Statistics, North Carolina State University\\ \smallskip}
\affil[*]{Corresponding author: laura.battagliola@epfl.ch}

\date{\today}

\begin{document}
\maketitle
\begin{abstract}
  \noindent

This article focuses on the study of lactating sows, where the main
interest is the influence of temperature, measured throughout
the day, on the lower quantiles of the daily feed intake. We outline a model framework and estimation methodology for quantile regression in scenarios with longitudinal data and functional
covariates. The quantile regression model uses a time-varying regression coefficient function to quantify the association between covariates and the quantile level of interest, and it includes subject-specific intercepts to incorporate within-subject dependence. Estimation relies on spline representations of the unknown coefficient functions, and can be carried out with existing software. We introduce bootstrap procedures for bias adjustment and computation of standard errors. Analysis of the lactation data indicates, among others, that the influence of temperature increases during the lactation period.

\medskip


\noindent


\end{abstract}

\newpage

\section{Introduction}

\noindent
This paper considers quantile regression for longitudinal data in the presence of functional covariates. It is motivated by data on the daily feed intake of lactating sows, where the aim is to study how temperature in the stable, or cell, during the day affects the feed intake, in particular for sows
that eat scarcely. This is of interest because poor nutrition in the lactation period may lead to health downsides, both for the sows and the piglets, and production inefficiency. 
Daily temperature is measured every fifth minute and is therefore naturally treated as a functional covariate,
and the study is longitudinal with both the feed intake and daily temperature profiles recorded for up to 21 days for each sow.

Quantile regression, first introduced by \cite{koenker1978regression}, is a well-established framework from statistics and econometrics. It is suitable when the analysis aims at describing and quantifying the association between covariates and quantiles of the distribution of the response variable. In particular, it allows to robustly target not only the central parts of the response distribution, but also the more extreme regions. For overviews, see the seminal monograph by \cite{koenker_2005} and for more recent developments, see \cite{koenker2017handbook}. 

Analyses of longitudinal data, including quantile regression, must account for the dependence between observations from the same subject in order to provide valid inference. 
A common approach is to include subject-specific effects in the model for the quantiles and use 
penalization, see for example  \cite{koenker2004quantile}, \cite{lamarche2010robust}, \cite{doi:10.1002/jae.2520}, \cite{GU2019}, and \cite{doi:10.1080/01621459.2020.1725521}. We adopt the same approach for this paper. 
Alternatives include \cite{RePEc:eee:econom:v:170:y:2012:i:1:p:76-91} and \cite{GALVAO201692}, who treated subject-specific parameters as fixed effects without penalization, and \cite{doi:10.1111/j.1368-423X.2011.00349.x}, who used a two-step procedure where subject-specific parameters are first estimated as fixed effects and then plugged in as offsets in a standard quantile regression  (see also \cite{RePEc:cfr:cefirw:w0249}). 

Quantile regression with functional covariates, similar to scalar-on-function mean regression, describes the association between a quantile of the response and a functional covariate using an inner product between the functional covariate and an unknown smooth coefficient function. As it is common in nonparametric regression, we approximate the coefficient function using a finite basis representation, and thus the infinite-dimensional estimation problem is converted to a finite-dimensional one. 
Pre-specified spline functions and eigenfunctions obtained from the spectral decomposition of the functional covariates' covariance operator are the most popular choices for selecting the basis functions, and they have both been used for quantile regression. For example, \cite{doi:10.1080/10485250500303015} and \cite{RePEc:hin:jnljps:3743762} used splines, whereas \cite{kato2012estimation},  \cite{https://doi.org/10.1111/j.1467-9868.2011.01008.x} and \cite{li2022inference} used eigenfunctions.
A related research area is additive quantile regression where the effect of a scalar covariate is modeled via a smooth function \citep{Fenske2013, greven2017general, geraci2019, doi:10.1080/01621459.2020.1725521}.

In this paper we consider functional quantile regression for scalar response and functional covariates, which are both observed repeatedly for many clusters or subjects.
To the best of our knowledge no papers in the literature are devoted to this situation.
We consider a set-up with longitudinal data and allow for the effect of the functional covariate on the quantile to evolve over observation time. 
We use penalized splines to handle the functional covariates and penalized cluster- or subject-specific intercepts to account for the dependence within clusters or subjects. The resulting model can be represented in a framework easily implementable using existing software from \citet{doi:10.1080/01621459.2020.1725521}. 
Moreover, we point out bias and variance issues of the estimators and propose adjustments obtained with bootstrap, using 
resampling techniques from \cite{battagliola2022} for bias adjustment and from  \cite{galvao2015bootstrap} for computation of standard errors. Altogether, our analysis gives new insight to the eating behavior of lactating sows, our ultimate goal. In particular, the analysis indicates that the association between temperature in the stable and the feed intake gets increasingly stronger after delivery.

The paper is structured as follows. The model framework and estimation methodology are described in Sections \ref{sec:framework} and \ref{sec:estimation}. We analyse the lactation data in Section \ref{sec:application} and summarise and discuss findings in Section \ref{sec:discussion}. The appendix provides details about the bootstrap procedures and the practical implementation. Finally, additional results from the application, results from simulation studies, and example code can be found in the supplementary materials.

\section{Framework}
\label{sec:framework}
\noindent
We consider data $\{(Y_{ij}, X_{ij}(\cdot), t_{ij})\}_{ij}$, with scalar responses $Y_{ij}$ and functional covariates $X_{ij}(\cdot)$ at time-points $t_{ij} \in \mathcal{T} \subset [0, \infty)$, where $i = 1, \ldots, N$ denotes clusters, and $j=1, \ldots, n_i$ denotes repeated measurements within cluster $i$. Observations from different clusters are assumed to be independent, but there may be within-cluster correlation. Covariates $X_{ij}(\cdot)$ are square-integrable functions on a closed interval $\mathcal{S} \subset \mathbb{R}$, i.e., $X_{ij}(\cdot)\in L^2(\mathcal{S})$. In practice they are often observed on a dense grid $\{s_1, \ldots, s_H\} \subset \mathcal{S}$ and possibly with measurement errors. 

We are concerned with quantile regression. Let $\tau \in (0,1)$ be a fixed quantile level, and assume that the $\tau$-quantile for the conditional distribution of the $j$-th observation $Y_{ij}$ from cluster $i$ given covariates $X_{ij}(\cdot)$ and $t_{ij}$ takes the form
\begin{equation}
\label{eq:target_quantile_model}
Q_{Y_{ij}|X_{ij}, u_i}^\tau(t_{ij}) = \alpha^\tau(t_{ij}) + \int_\mathcal{S} \beta^\tau(t_{ij},s) X_{ij}(s) ds + u_i,
\end{equation}
where $u_i$ (dependence of $\tau$ suppressed in notation) specifies a cluster-specific level. The target parameters of the analysis are the intercept functions $\alpha^\tau(\cdot)$ and the regression coeffient function $\beta^\tau (\cdot, \cdot)$ which are both assumed to be common for all clusters. 
In particular, the functional covariate affects the $\tau$-quantile in the same way for all clusters. 
Without further restrictions, $\alpha^\tau(\cdot)$ is identifiable up to an additive constant, and $\beta^\tau(t, \cdot)$ is identifiable up to an additive component in the orthogonal complement of the vector space spanned by the functional covariates $X_{ij}(\cdot)$.

The notation in (\ref{eq:target_quantile_model}) reflects that we think of data as emerging from a two-step process: $u_i$ is a sample of i.i.d.\ random variables with mean zero, and $Y_{ij}$'s are then generated independently from a model with $\tau$-quantile \eqref{eq:target_quantile_model}. The restriction $\mathbb{E}[u_i]=0$ ensures full (asymptotic) identifiability of $\alpha^\tau(\cdot)$. We emphasize that \eqref{eq:target_quantile_model} does not specify the full conditional distribution of $Y_{ij}$ given $\{X_{ij}(\cdot), t_{ij}\}$ in cluster $i$, only its $\tau$-quantile, and we suggest to use it for one or a few quantile levels of particular interest.

With the two-step data generating process, we may also consider the $\tau$-quantile of the conditional distribution of $Y_{ij}$ given $\{X_{ij}(\cdot),t_{ij}\}$ \emph{marginally over all clusters}.
The association between this implied marginal $\tau$-quantile and $X_{ij}$ may not take a form similar to that of \eqref{eq:target_quantile_model}. Ignoring the cluster-specific parameters, i.e., fitting the quantile regression model \eqref{eq:target_quantile_model} with all $u_i=0$, 
would therefore not target $\alpha^\tau(\cdot)$ and $\beta^\tau(t, \cdot)$. This is an important difference compared to the associated mean regression mixed-effects model where the conditional and marginal means would be described by the same coefficient function, such that an analysis based on the marginal model would lead to reliable estimates (but possibly wrong inference). See the supplementary materials and \citet{battagliola2022} for further considerations on marginal versus conditional models and analyses in quantile mixed-effects models.

\section{Estimation methodology}
\label{sec:estimation}
\noindent
Two main challenges arise for the estimation of the model \eqref{eq:target_quantile_model} compared to classical quantile regression for independent data with scalar covariates: how to represent the longitudinal and longitudinal functional coefficients $\alpha^\tau(\cdot)$ and $\beta^\tau(\cdot,\cdot)$, and how to handle the cluster-specific intercepts $u_i$.

\subsection{Representation of the functional coefficient and smooth intercept}
\label{sec:estimation_spline_rep}
\noindent
Firstly, we assume that $t\mapsto \alpha^\tau(t)$ and $(s,t)\mapsto \beta^\tau(s,t)$ depend smoothly on time. This allows us to use tools from additive models \citep{wood2017generalized} and hence approximate the functions along the $t$-coordinate with some basis functions $\{\psi_l(\cdot)\}_{l=1}^L$. For simplicity, we choose to use the same basis functions for both coefficient functions. 
Secondly, in the functional regression literature it is also common to have a finite-dimensional representation of functional coefficients in the $s$-coordinate. Let $\{\varphi_d(\cdot)\}_{d=1}^D$ be basis functions for that purpose, and represent $\beta^\tau(\cdot,\cdot)$ as  a tensor product smooth with different bases for the two coordinates. To be specific, we write
\begin{equation}
\label{eq:coefficients_approx}
\begin{aligned}
    \alpha^\tau(t) & \approx \sum_{l=1}^L a^\tau_l\psi_l(t),\\
    \beta^\tau(s,t) & \approx \sum_{l=1}^L \sum_{d=1}^D \delta^\tau_{dl}\psi_l(t) \varphi_d(s),
\end{aligned}
\end{equation}
where $a^\tau_l$'s and $\delta^\tau_{dl}$'s are unknown coefficients. In the application we use cubic spline bases; in the following we describe the methodology for general penalized splines. 

With the representations in \eqref{eq:coefficients_approx}, the integral in \eqref{eq:target_quantile_model} is approximated with
$$\int_\mathcal{S} \beta^\tau(t,s) X_{ij}(s) ds \approx \sum_{l=1}^L \sum_{d=1}^D \delta^\tau_{dl}\psi_l(t) \int_\mathcal{S}  \varphi_d(s)X_{ij}(s) ds,$$
and we can work with finite-dimensional version of model \eqref{eq:target_quantile_model}, namely
\begin{equation}
\label{eq:te_functional_term}
	Q_{Y_{ij}|X_{ij}, u_i}^\tau(t_{ij})  = \sum_{l=1}^L a^\tau_l \psi_l(t_{ij}) +\sum_{l=1}^L \sum_{d=1}^D  \delta^\tau_{dl} \psi_l(t_{ij}) \xi_{d,ij} + u_i.
\end{equation}
Here, $\xi_{d,ij} = \int_\mathcal{S}  \varphi_d(s)X_{ij}(s) ds$, and the coefficients $\{a^\tau_l\}_{l}$ and $\{\delta^\tau_{dl}\}_{dl}$ and the subject-specific intercepts $\{u_i\}_i$ are the unknown parameters. In practice, the integrals $\xi_{d,ij}$ are approximated with Riemann sums.

\subsection{Estimation of coefficients and random intercepts}
\label{subsec:est}
\noindent
In standard quantile regression, parameters are estimated by minimizing an empirical loss, $\sum_{i=1}^N  \sum_{j=1}^{n_i} l_\tau(Y_{ij}-Q^\tau_{ij})$, where $Q_{ij}^\tau$ is short for the level $\tau$ quantile for observation $j$ of cluster $i$ and depends on the model parameters, and $l_\tau$ is an appropriate loss function, typically the check loss function $v\mapsto v (\tau - \mathbbm{1}_{(v <0)})$ \citep{koenker1978regression}.
The method proposed by \citet{doi:10.1080/01621459.2020.1725521}, named "QGAM", offers a flexible framework to model longitudinal quantile regression with scalar covariates and is implemented in an
accompanying R package \texttt{qgam}. QGAM uses a smooth approximation of the check loss function in order to make it differentiable so common computational optimizers like the Newton method can be used for the minimization problem. 

We adapt QGAM to functional covariates. We penalize the subject-specific intercepts $\{u_i\}_i$ and coefficients $\{a^\tau_l\}_{l=1}^L$ and $\{\delta^\tau_{dl}\}_{d=1,l=1}^{D,L}$, as it is common in the additive mixed-effects models literature. In particular, for the former an $\ell_2$-penalty is added to the loss function, while for the latter penalty terms accounting for wiggliness in the $s$- and $t$-directions are added \citep[Chapter 5]{wood2017generalized}. The tuning parameters defining the degree of penalization are selected as part of the procesdure as implemented in the \texttt{qgam} package, see \cite{doi:10.1080/01621459.2020.1725521} for details. In the following, we refer to the extension of QGAM to functional covariates as "fQGAM".

\subsection{Covariates observed with noise}
\label{sec:noisy_covariates}
\noindent
In the previous sections the covariate functions were assumed to be observed densely and without measurement noise. 
Now, consider the more realistic situation with measurement noise and possibly sparse sampling, and denote by $W_{ij,h}$ the observations corresponding to points $s_h$, i.e., 
\begin{equation}
W_{ij,h} = X_{ij}(s_{h}) + \epsilon_{ij,h} \hspace{5 mm}h=1,\ldots,H,
\label{eq:W}
\end{equation}
\color{black}
where $\{\epsilon_{ij,h}\}_{ijh}$ are  iid.\ random variables with mean zero, and mutually independent of the underlying functions. We propose to carry out a preliminary smoothing step and proceed with the analysis from Section~\ref{subsec:est} with the unobserved values $X_{ij}(s)$ replaced by their 
fitted/predicted values $\hat{X}_{ij}(s)$.

There are many smoothing techniques available for functional data, e.g., kernel-based methods, smoothing splines, and smoothing with data-driven bases, see for example \cite{ramsay2005functional}.
We choose to represent $\{\hat{X}_{ij}(\cdot)\}_{ij}$ with eigenfunctions arising from Functional Principal Component Analysis (FPCA), by assuming independence over both $i$ and $j$. The number of eigenfunctions should be large enough to capture the primary modes of variation of $\{W_{ij,h}\}_{ijh}$, but small enough to get smooth reconstructed functions. The choice is usually based on a preset Percentage of Variance Explained (PVE).
Several implementations of FPCA are available depending on the sampling pattern of the functional data (dense or sparse, same or different sampling locations, missing values), see e.g. \citet{Yao2003}, \cite{xiao2018}, and references therein. 
We use the fast covariance estimation (FACE) method from \cite{xiao2016fast} in this work, ignoring potential dependence among functions. This is not inappropriate
functions (see e.g. \cite{Goldsmith2012}).
Approaches that account for dependence within subject or cluster have been discussed by \cite{
greven2010, chen2012modeling, park2015, koner2023second}, to name a few.

\subsection{Bootstrap procedures for variance assessment and bias adjustment}
\label{sec:boot}

\noindent
In the sow data application we are mainly interested in estimation and inference for quantiles 
and the differences between quantiles at specified directions of the functional covariate. 
It is known from the literature on quantile regression for longitudinal data with scalar covariates, that estimators may be biased and that it is difficult to properly assess the sampling variability of the estimators without resampling methods \citep{RePEc:eee:econom:v:170:y:2012:i:1:p:76-91, galvao2015bootstrap, battagliola2022}. 
We have seen in simulation studies (available in the supplementary materials) that the problems persist when covariates are functional, and we propose to use bootstrap strategies for variance estimation and bias adjustment.

Recall the quantile model \eqref{eq:target_quantile_model} with repeated measurements of functional covariates and responses for each subject. 
Our target parameters are described as follows. 
Consider a fixed time point $t$, a function $X(\cdot)\in L^2(\mathcal{S})$ and response $Y$. The corresponding \textit{linear predictor at level $\tau$}  is
$$Q_{Y|X, 0}^\tau(t)  =\alpha^\tau(t)+\int_\mathcal{S} \beta^\tau(s,t)X(s)\, ds$$
and is interpreted as the $\tau$-quantile for a typical subject (with $u=0$). The function $X(\cdot)$ may or may not be one of the functions in the dataset.
Furthermore, consider two functional covariates $X_A(\cdot),X_B(\cdot)\in L^2(\mathcal{S})$ with pointwise difference, $\Delta X(s) = X_A(s) - X_B(s)$. For a fixed cluster, i.e.\ a fixed $u$ and a fixed measurement time $t$, the corresponding \textit{difference in the $\tau$-quantile} is 
\begin{equation}
D^\tau(t) = Q_{Y|X_A, u}^\tau(t) -  Q_{Y|X_B, u}^\tau(t)=
\int_{\mathcal{S}} \beta^\tau (s,t) \Delta X(s)\, ds,
\label{eq:diff_quantiles}
\end{equation}
so $D^\tau(t)$ is the difference in quantile for a fixed subject when $X(\cdot)$ is changed in direction $\Delta X(\cdot)$. In the following we refer to $Q_{Y|X, 0}^\tau(t)$ or $D^\tau(t)$ as targets $\theta$ of interest and let $\hat \theta$ denote the corresponding estimate calculated from estimates of the coefficients $\beta^\tau(\cdot,\cdot)$ and  $\alpha^\tau(\cdot)$ in the representation \eqref{eq:coefficients_approx}.

First, consider estimation of $\text{var}(\hat\theta)$. The estimated model coefficients, $\{\hat a^\tau_l\}_l$ and $\{\hat\delta^\tau_{dl}\}_{dl}$, from \texttt{qgam} are accompanied with a variance-covariance matrix 
which can be used for computation of a standard error for the estimator $\hat \theta$. 
We refer to these standard errors as \textit{model-based standard errors}. However, penalization of random effects is likely to cause underestimation of the true sampling variation of $\hat\theta$.
As \cite{galvao2015bootstrap}, we resample complete subject data with replacement to compute standard errors. 
Specifically, let $\tilde \theta_1,\ldots,\tilde \theta_B$ be estimates when the estimation procedure is applied to $B$ bootstrap datasets, and use $\text{sd}_{\text{boot}}(\hat \theta) =  
\text{sd}(\tilde \theta_1,\ldots,\tilde \theta_B)$ as an estimate for $\sqrt{\text{var}(\hat\theta)}$. Details of the sampling procedure can be found in the appendix.

Second, consider estimation of $\text{bias}(\hat\theta)$. 
As documented by \cite{battagliola2022}, bias can occur even for large samples, caused by a combination of the incidental parameter problem (the number of parameters increase with sample size, \citet{neyman1948consistent, Lancaster2000}), non-linearity of quantiles, and  penalization of the subject-specific intercepts.
Block resampling cannot be used for bias adjustment because the target parameter of interest is not computable under the bootstrap distribution; see also \cite{Karlsson2009} who obtained little or no effect in an attempt to adjust for bias in a nonlinear quantile regression for longitudinal data.
Instead, we propose to combine standard resampling of estimated random effects with wild bootstrap of residual terms, using the technique developed by \cite{battagliola2022}. The purpose is to generate bootstrap datasets 
under a distribution where the true value of the target parameter coincides with $\hat \theta$ (the estimate obtained from the observed data), such that the bias can be estimated from the bootstrap estimates. 
Specifically, let $\tilde \theta_1,\ldots,\tilde \theta_B$ be estimated values of $\theta$ for $B$ bootstrap datasets, then bias is estimated as $\text{bias}_{\text{boot}}(\hat \theta) =  \frac{1}{B}\sum_{b=1}^B(\tilde \theta_b - \hat \theta)$. We explain the sampling procedure in more detail in the appendix.

The block resampling method and the sampling method suggested by \cite{battagliola2022} differ in several ways.
While the former is completely non-parametric, the latter relies on the model. 
Another important difference is that the covariate functions are resampled (together with the responses) by the 
block resampling method, but kept exactly as in the dataset in the approach of \cite{battagliola2022}.
As a consequence, the procedure based on wild bootstrap would underestimate the variance of the estimator $\hat \theta$. 
Our suggested solution is to combine the estimated bias and estimated standard deviation from the two bootstrap sampling methods, respectively, to construct confidence intervals for the target $\theta$. If the distribution of $\hat \theta$ is well approximated by a normal distribution, and standard errors and bias are estimated as described above, it is natural to define approximate $1-\alpha$ confidence intervals as
\begin{equation}
\label{eq:bootCI}
    \hat \theta- \text{bias}_\text{boot}(\hat \theta) \ \pm \ q_{1-\alpha/2} \ \text{sd}_\text{boot} (\hat \theta).
  \end{equation}
  
\cite{battagliola2022} demonstrated in a wide variety of simulation settings with clustered data and scalar covariates that bias was greatly reduced or removed, with the above bootstrap sampling process combining resampled cluster-specific intercepts and wild bootstrap for the residuals, and \cite{galvao2015bootstrap} demonstrated that sampling variation of estimators is measured appropriately with block resampling. 
We acknowledge that investigations of the coverage properties of the confidence intervals in the current setting would be of interest, but leave it for future research and focus on the application below.

\section{Lactating sows' feed intake}
\label{sec:application}
\noindent
Our ultimate goal is to study the impact of thermal conditions on the daily food intake for lactating sows, taking into account the progression of food intake over lactation days. 
(compromised reproductive system) as well as for its litter. In particular, low food intake may lead to
increased body weight loss and reduced milk production, implying slower and poorer weight gain of the litter 
(see for instance \cite{rosero2016essential}, \cite{bloemhof2013effect}, \cite{renaudeau2001effects}, \cite{ncr1999effect}, \cite{renaudeau2001effects1}).
In this paper we focus on understanding the association between low quantile levels of feed intake and the hourly temperature and how this association varies over time.

\subsection{Description and preprocessing of data}

\noindent
The data comes from a commercial research unit in Oklahoma, where 480 sows were monitored from July to October 2013. The animals were divided into 21 groups and then assigned to cells, where they were kept under observations during the lactation period for up to 21 days.
For each sow at each lactation day, the food intake (in kg) is available, as well as the
cell temperature (in \textdegree C), measured every five minutes for 24 hours from 2.00 pm to 1.59 pm the following day. Moreover, the parity of each sow is registered, i.e., the number of pregnancies the animal had before the current one. We will consider parity as a measure of age: a sow is ``young'' if it is at its first pregnancy and it is ``old'' otherwise. Previous studies have shown that younger and older sows behave differently \citep{doi:10.1111/rssc.12376}, so we analyze data from young and older sows separately.
There are 475 sows in total with 237 young sows and 238 old ones, respectively.

\begin{figure}[H]
\center
\includegraphics[width=12cm]{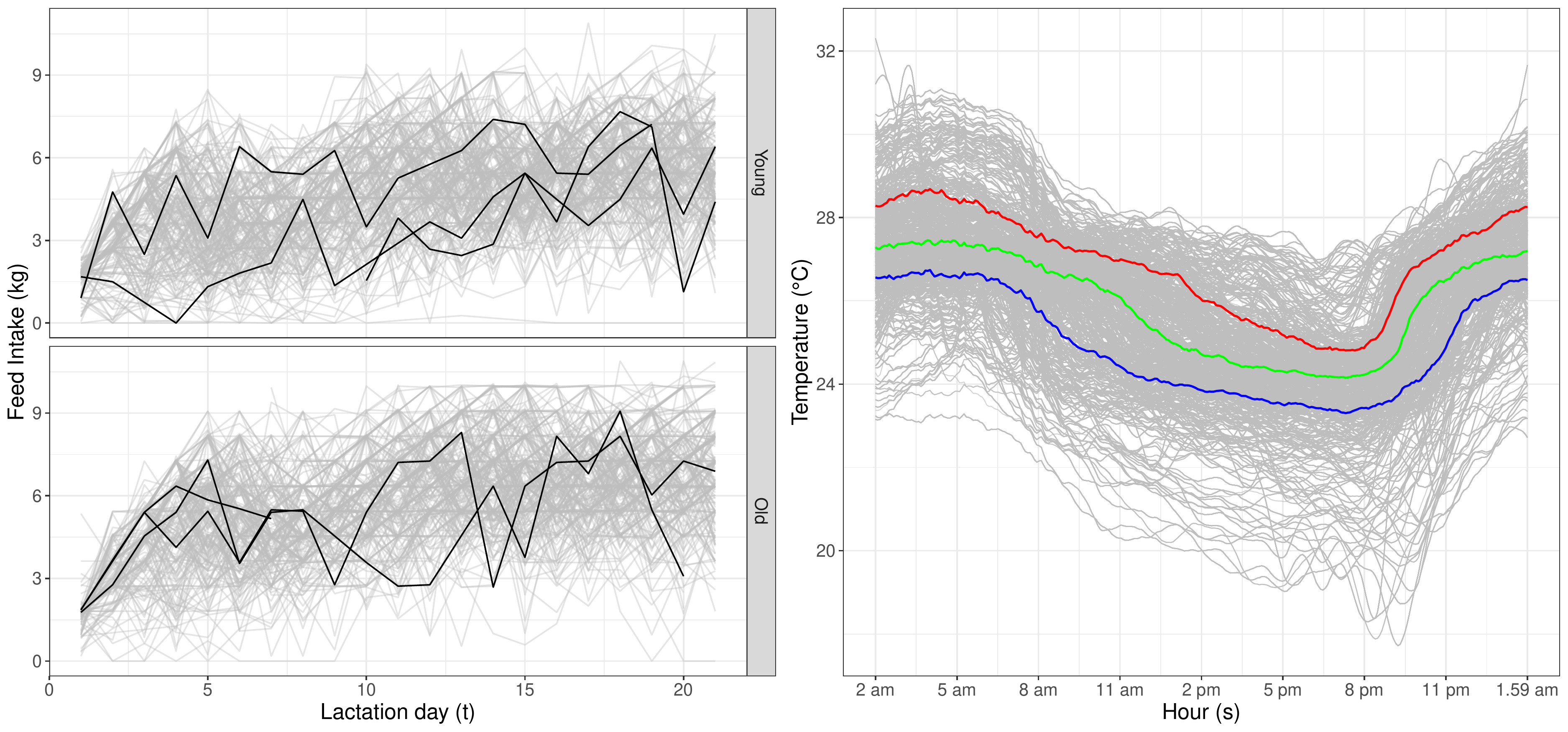}
	\caption{{The lactation data. Left: daily feed intake profiles over lactation days of young sows (upper panel) and of old sows (lower panel) with three randomly selected profiles (black) in each group. 
 For some sows data are only available in a subset of the lactation period. Right: smoothed temperature curves (grey), as well as the pointwise temperature quantiles curves at quantile levels 20\% (blue), 50\% (green) and 80\% (red) based on the whole dataset.}}
\label{fig:application_temperature}
\end{figure}

The data are illustrated in Figure~\ref{fig:application_temperature}. Feed intake profiles are plotted in the left panel with profiles from three randomly selected sows from each age group highlighted. Although there is large within-sow variation over lactation days, it is also clear that some sows tend be have low (or high) feed intake throughout, calling for a {subject-specific component in the model}.  
In a preprocessing step we smoothed the temperature curves with FACE (see Section \ref{sec:noisy_covariates}), using a PVE of 99.99\% so that most of the features of the curves are maintained. The reconstructed daily temperature trajectories are used in the analysis.

\cite{doi:10.1111/rssc.12376} used a longitudinal dynamic functional regression framework for mean regression for the same data with emphasis on prediction of response trajectories. \cite{RePEc:hin:jnljps:3743762} carried out separate quantile regression analyses for a derived variable at three selected lactation days. 
For each day separately, the cumulative distribution function (CDF) was first estimated and then inverted to estimate quantiles of interest. 
In contrast, we carry out quantile regression for all lactation days simultaneously using the model framework and estimation method introduced in Sections \ref{sec:framework} and \ref{sec:estimation}, and our analysis provides estimates and confidence bands for the temperature effect on quantiles of feed intake. In particular, we consider the estimated quantiles of the feed intake when the daily temperature corresponds to the pointwise 20\% and 80\% quantiles of the smoothed temperature curves. These two temperature profiles, denoted $\text{Temp}_{20}(\cdot)$ and $\text{Temp}_{80}(\cdot)$, respectively, and shown in Figure \ref{fig:application_temperature} in blue and red, are the most extreme temperature quantiles considered by \cite{RePEc:hin:jnljps:3743762}. The pointwise median temperature curves
is also plotted in Figure \ref{fig:application_temperature} (green).

\subsection{Estimated quantiles of feed intake}
\label{sec:application_predQ0}
\noindent
Denote the observed data by $\{(\text{FI}_{ij}, \text{Temp}_{ij}(\cdot), t_{ij})\}_{ij}$. For each sow $i=1,\ldots,N$ ($N=237$ or $N=238$) and repeated measurement $j=1,\ldots,n_i$ ($n_i$ ranging from {7} to {21}), 
$\text{FI}_{ij}$ refers to the daily feed intake expressed in kg, $\text{Temp}_{ij}(\cdot)$ to the smoothed temperature function in \textdegree C recorded over a day and $t_{ij}$ to the lactation day. We allow for a subject-specific intercept $u_i$ to account for the correlation of observations from the same sow.
For each age group, we consider the model 
\begin{equation}
\label{eq:application_model_st}
Q_{\text{FI}_{ij}|\text{Temp}_{ij},  u_i}^\tau(t_{ij}) =  \alpha^\tau(t_{ij}) + \int_\mathcal{S} \beta^\tau(s,t_{ij})\text{Temp}_{ij}(s)ds + u_i, \quad i=1,\ldots,N, \ j=1,\ldots,n_i
\end{equation}
where $\mathcal{S}$ represents a whole day from 2.00 pm to 1.59 pm. We approximate the smooth intercept
$\alpha^\tau(\cdot)$ using ten cubic splines 
and the coefficient function $\beta^\tau(\cdot,\cdot)$ using a tensor product of ten cubic splines in both directions, with cyclic splines for the $s$-direction.

\begin{figure}[H]
\center
\includegraphics[width=12cm]{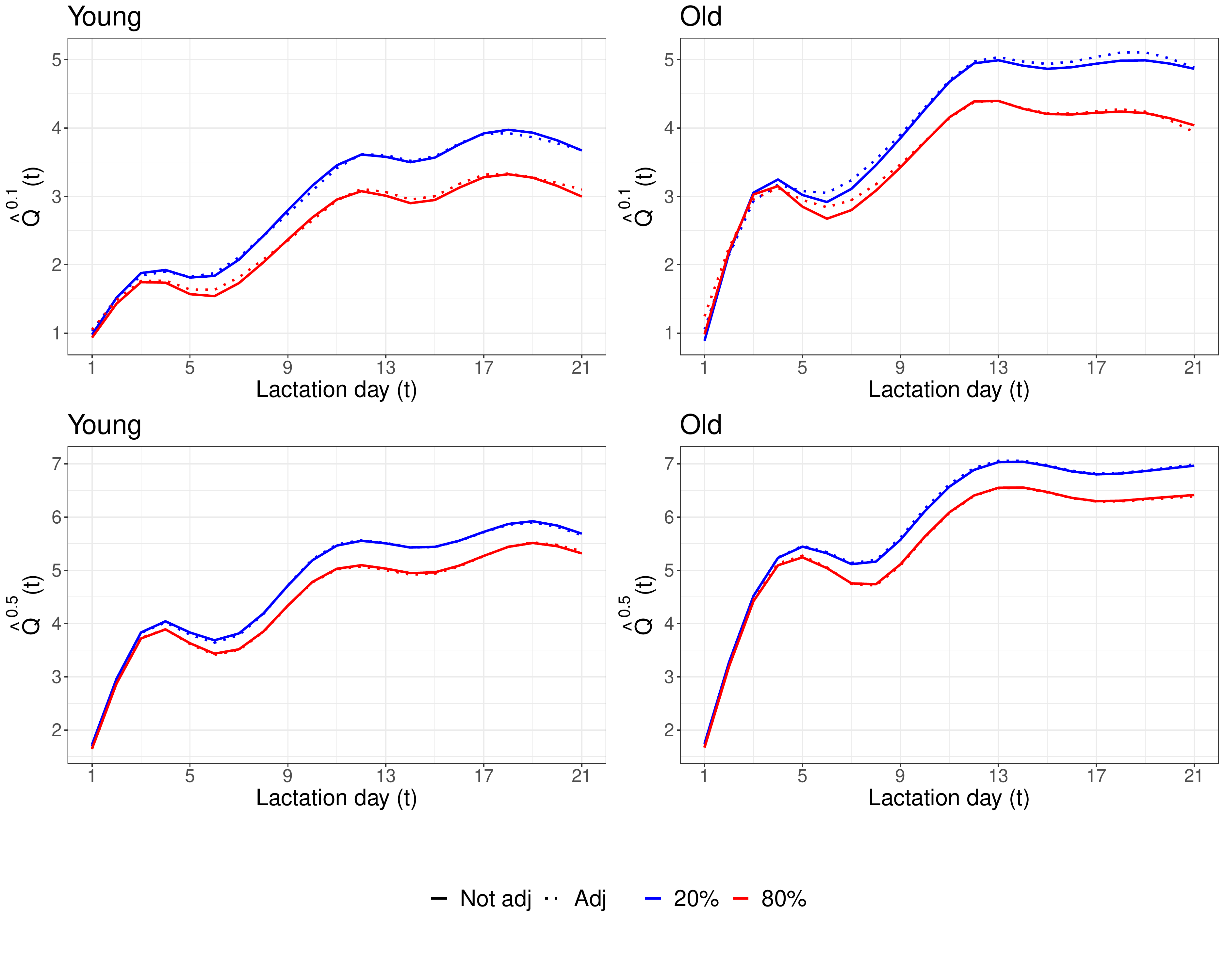}
	\caption{
Predicted quantiles corresponding to the 20\% and 80\% pointwise temperature profiles. Bootstrap-adjusted estimates are shown with dotted curves. The left column refers to sows at their first pregnancy, while right one refers to the older sows. Results at quantile levels $\tau=0.1$ and $\tau=0.5$ are shown in the top and bottom row, respectively. Notice that predicted quantiles at different quantile levels are plotted on different scales.
}
\label{fig:application_pred_model_comparison}
\end{figure}

Figure \ref{fig:application_pred_model_comparison} shows estimated quantile profiles for young/old sows (left/right), at quantile levels 0.1/0.5 (top/bottom), and for the pointwise 20\% and 80\% temperature curves $\text{Temp}_{20}(\cdot)$ and $\text{Temp}_{80}(\cdot)$ (colours as above). More specifically, the graphs show
$$
\hat Q^\tau_{Temp_{20},0}(t) = \hat\alpha^\tau(t) + \int_\mathcal{S} \text{Temp}_{20}(s)\hat\beta^\tau(s,t) ds
,\quad
\hat Q^\tau_{Temp_{80},0}(t) = \hat\alpha^\tau(t) + \int_\mathcal{S} \text{Temp}_{80}(s)\hat\beta^\tau(s,t) ds,
$$
plotted over $t$, for each age group and for $\tau=0.1,0.5$. Notice that no random effects are included in the predictions such that their interpretation is for a ``typical sow''. 

The solid curves are estimated profiles, and we see a clear distinction between low (blue) and high (red) temperatures, at least from around lactation day five. High temperatures negatively influence the appetite of the sows---they tend to eat more in cooler conditions---and this difference increases over time, particularly at the 0.1 level.
%
The group of young and old sows have similar quantiles of feed intake at the very beginning of their lactation period, followed by a steep increase up to around lactation day 5, a short period with constant feed intake, and a final increase up to a stable plateau. However, estimated increments appear to be smaller for young sows.
The dashed curves show the bias-adjusted estimates. Although the bias adjustment is hardly visible, it is actually significantly different from zero at many instances at a 5\% significance level (based on pointwise one-sample $t$-tests on the estimates from the bootstrap data). 

\begin{figure}[H]
\center
\includegraphics[width=12cm]{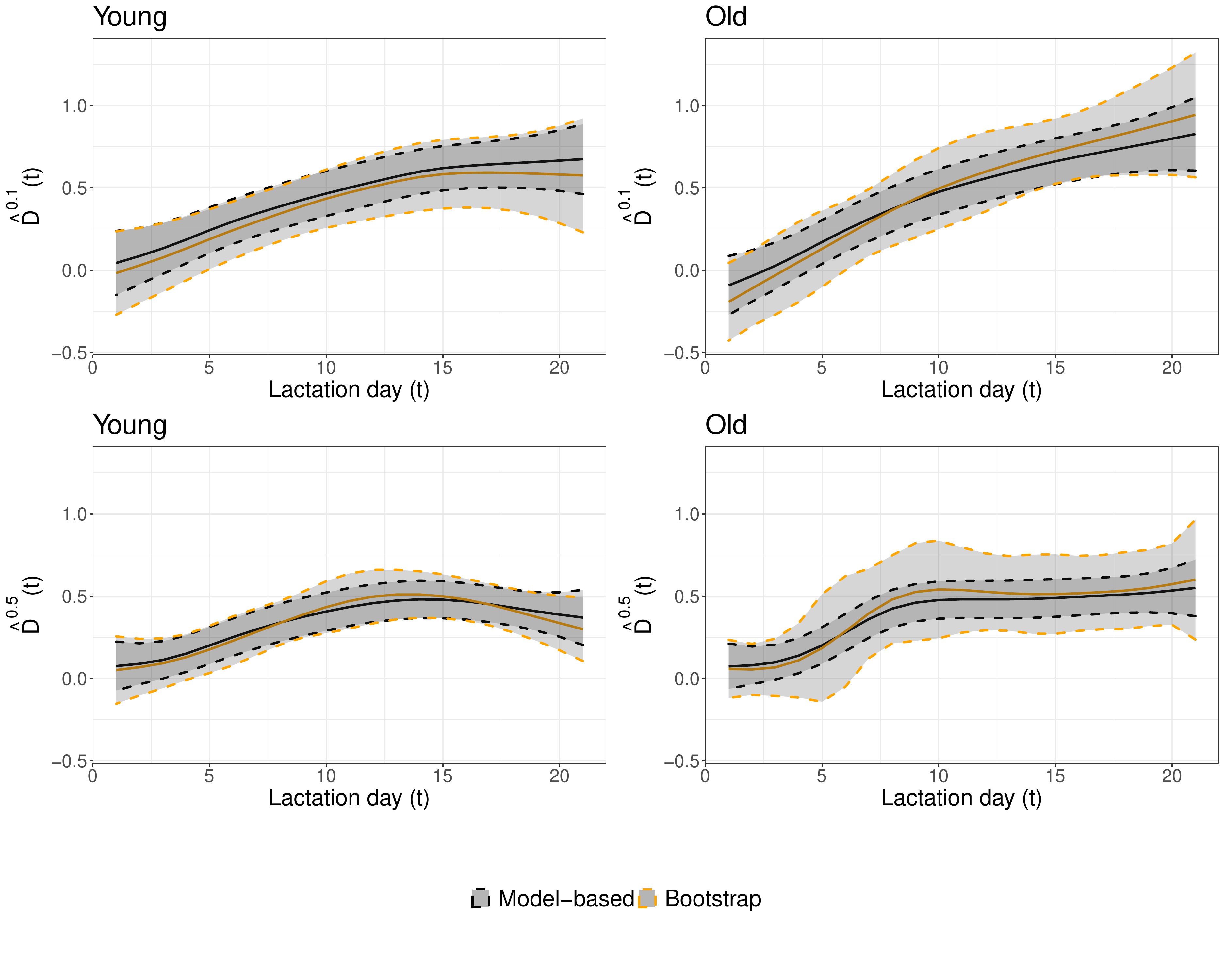}
\caption{Estimated differences in quantiles between the pointwise 20\% and 80\% temperature curves, both without (solid black) and with (solid orange) bias adjustment. The corresponding pointwise confidence intervals,
  based on the model solely or on bootstrap, are illustrated with dashed curves. The left column refers to sows at their first pregnancy, while the right column refers to the older sows. Results at levels $\tau=0,1$ and $\tau=0.5$ are shown in the top and bottom row, respectively.}
\label{fig:application_pred_model_comparison_diff}
\end{figure}

In order to illustrate the estimated temperature effects
more clearly, Figure \ref{fig:application_pred_model_comparison_diff} shows the difference between the estimated feed intake quantile profiles for low and high temperatures, {i.e.\ ${\hat D}^\tau(t) =  \hat Q^\tau_{Temp_{20},0}(t)- \hat Q^\tau_{Temp_{80},0}(t)$}.
The black curves and confidence bands show the estimates without adjustment and the corresponding model-based 95\% pointwise confidence interval based on the variance-covariance matrix extracted from the model fit, and the orange curves and confidence bands show the bias-adjusted estimates and confidence bands obtained by bootstrap, as in equation \eqref{eq:bootCI}. We used 100 bootstrap samples for bias adjustment as well as for computation of confidence intervals, cf.\ Section \ref{sec:boot}. 
Bias adjustment is most noticeable for young sows at the 0.1 quantile, 
and the bootstrap generated confidence bands are always wider than the model-based ones. Simulation results have indicated that the model-based standard errors underestimate the actual variation (see the supplementary materials), so we prefer the bootstrap generated confidence intervals. For completeness, the profiles obtained from bootstrap datasets are shown in Figure S6 in the supplementary materials.

Irrespective of the method used for construction of confidence bands and of the bias adjustment, the overall conclusion is the same:
No temperature effect is found early in the lactation period (up to around day five), but at later days quantiles of feed intake are negatively affected by high temperature, both at 0.1 and 0.5 quantile levels.
In general, the influence of temperature on the quantiles becomes more prominent along the lactation period, but there are certain differences between age groups and between quantile levels. At quantile level 0.1 the difference in estimates has an increasing trend along lactation days for both groups of sows, while at the median the difference in estimates reaches a maximum of approximately 0.5 around lactation day 10--13 and then flattens out. This might indicate that the sows that eat less are those particularly sensible to the environmental temperature.

\subsection{Comparison with simpler models}
\label{sec:simplermodels}
\noindent
Now, let us turn to a comparison of the model \eqref{eq:application_model_st}, with three simpler alternatives. The first modification has $\beta(s,t) \equiv \beta_A(s)$ such that the temperature curves still have functional effects, but with same effect across lactation days; this would correspond to profiles of differences in Figure~\ref{fig:application_pred_model_comparison_diff} being constant. The second modification has $\beta(s,t) \equiv\beta_B(t)$. Then the model \eqref{eq:application_model_st} becomes
\begin{equation}
\label{eq:application_model_t}
Q_{\text{FI}_{ij}|\text{Temp}_{ij}, u_i}^\tau(t_{ij}) = \alpha^\tau(t_{ij}) + \beta_B^\tau (t_{ij}) \int_{\mathcal{S}} \text{Temp}_{ij}(s)\, ds + u_i
\end{equation}
such that the quantile depends on the temperature curve only through its integral or, equivalently, the average temperature over the day, and it is no longer a \textit{functional} quantile regression model. The third modification combines the two previous sub-models; it has $\beta(s,t) \equiv \beta_C$, such that 
\begin{equation}
\label{eq:application_model_const}
Q_{\text{FI}_{ij}|\text{Temp}_{ij}, u_i}^\tau(t_{ij}) = \alpha^\tau(t_{ij}) + \beta_C^\tau \int_{\mathcal{S}} \text{Temp}_{ij}(s)\, ds + u_i
\end{equation}
and the temperature effect is the same across days and depends on the average temperature over the day only.

We measure goodness-of-fit with the AIC values based on the log-likelihood corresponding to the Extended log-F (ELF) distribution \citep{doi:10.1080/01621459.2020.1725521} and the effective degrees of freedom (EDF) known from additive models \citep{wood2017generalized}. 
The EDF is partitioned into two parts: the effective degrees of freedom for the smooth coefficients, denoted EDF$_{\alpha,\beta}$, and the degrees of freedom corresponding to the subject-specific intercepts, denoted EDF$_u$.

The results are displayed in Table \ref{table:AIC}. For both groups the AIC values are notably smaller for the most complex model than for its competitors for quantile level $\tau=0.1$, while the values are closer among the models at the median. This indicates that it is particularly important to allow for time-varying coefficients and functional effects at lower quantiles. At level $\tau=0.5$ the most complex model is still selected for old sows, but for younger animals the smallest AIC is the one from model \eqref{eq:application_model_t}. Furthermore, in all cases, the AIC values from the model with $\beta_B(t)$ are smaller than the AIC value from the model with $\beta_A(s)$, indicating that it is more important to account for the temperature variation in the development along the lactation period than over the day.

For both groups and at both quantile levels, EDF$_{\alpha,\beta}$ is the highest for the model \eqref{eq:application_model_st}, as expected, since it describes variation in both the $s$ and $t$ direction. 
Both EDF$_{\alpha,\beta}$ and EDF$_u$ are larger when estimation is carried out at the median rather than at the 10\% level; most likely because there is more information in the data to estimate the median, which in turn allows for higher flexibility. Finally, EDF$_u$ is always between 188 and 207, and thus smaller than 237 and 238, the number of young and old sows, respectively, so random effects are penalized to some degree. 

\begin{table}[H]
\small
\centerline{
\renewcommand{\arraystretch}{1.5}
\begin{tabular}{rc|ccc|ccc|ccc|ccc}
& & \multicolumn{3}{c}{$\beta(s,t)$} & \multicolumn{3}{c}{$\beta_A(s)$} & \multicolumn{3}{c}{$\beta_B(t)$} & \multicolumn{3}{c}{$\beta_C$} \\
  \cline{2-14}
& $\tau$ & AIC & EDF$_{\alpha,\beta}$ & EDF$_{u}$ & AIC & EDF$_{\alpha,\beta}$ & EDF$_{u}$ & AIC & EDF$_{\alpha,\beta}$ & EDF$_{u}$ & AIC & EDF$_{\alpha,\beta}$ & EDF$_{u}$\\ 
    \cline{2-14}
   \parbox[t]{2mm}{\multirow{2}{*}{\rotatebox[origin=c]{90}{Young}}}& $0.1$ & \emph{18527} & 19 & 191 &18769 &13& 188 &18724 & 13& 189& 18790 & 10& 188\\ 
& $0.5$ & 15937 &  19 & 206& 15947 & 14 & 206 & \emph{15934} & 14 & 206 & 15946 & 11 & 206\\ 
    \cline{2-14}
\parbox[t]{2mm}{\multirow{2}{*}{\rotatebox[origin=c]{90}{Old}}} & $0.1$ & \emph{18483} & 24 & 201 &18769&15& 198 &18633 & 13& 199& 18802 & 11& 198\\ 
& $0.5$ & \emph{16044} &  26 & 206& 16048& 15 & 207 & 16047 & 14 & 206 & 16060 & 11 & 207\\ 
    \cline{2-14}
\end{tabular}
\normalsize
}
\caption{AIC and sum of effective degrees of freedom for young and old animals when adopting model \eqref{eq:application_model_st} (first column), model \eqref{eq:application_model_st} with $\beta(s,t) = \beta_A(s)$ (second column), model \eqref{eq:application_model_t} (third column) and model \eqref{eq:application_model_const} (fourth column). The smallest values of AIC are emphasized in each row.}
\label{table:AIC}
\end{table}

\subsection{Estimated effect of temperature} 

\begin{figure}[H]
\center

\includegraphics[width=12cm]{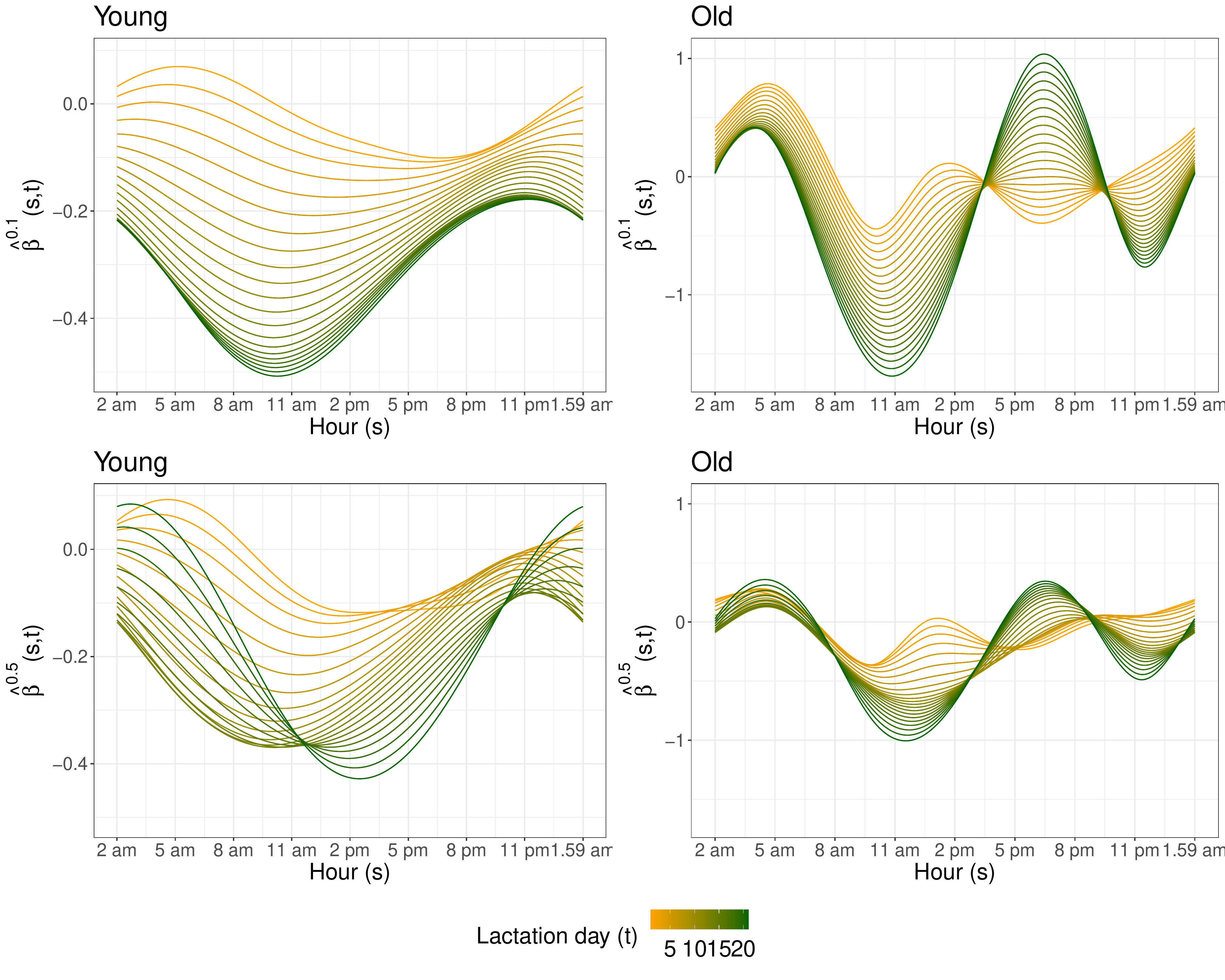}
	\caption{Illustration of the estimated coefficient function $\hat{\beta}^\tau(\cdot,\cdot)$ at level $\tau=0.1$ (top row) and $\tau=0.5$ (bottom row), for both young (left column) and older sows (right column). Curves show $s\mapsto \hat{\beta}^\tau(s,t)$ for each lactation days $t$, varying in color.}
	\label{fig:coef_splines_betast}
\end{figure}

\noindent
Finally, we turn the attention to the estimated coefficient function $\hat\beta^\tau(\cdot,\cdot)$
in model \eqref{eq:application_model_st}. 
It is important to keep in mind that the estimated functional coefficient is only identifiable up to elements belonging to the orthogonal complement of the space spanned by the basis functions used in the finite dimensional representation of the functional covariates. Thus the interpretation of $\hat\beta^\tau(\cdot,\cdot)$ must be taken with caution. 
Figure \ref{fig:coef_splines_betast} shows $\hat\beta^\tau(\cdot,\cdot)$ for each age group and at quantile levels 0.1 and 0.5, respectively. In each panel, $s\mapsto \hat\beta^\tau(s,t)$ is plotted for $t$ fixed at each lactation day, on a colour scale that ranges from orange to green as the longitudinal time $t$ goes by. Recall that, by construction, the development in both $s$ and $t$ direction is smooth, and the functions are cyclic over day.

The estimated coefficient functions are predominantly negative, corresponding to an overall negative effect of temperature, cf.\ Figure \ref{fig:application_temperature}. With the risk of overinterpretation, we see that the impact of temperature on feed intake is most prominent in the morning hours (from about 8 am to about 12, a bit later at late lactation days for young sows at the median). This is also the time of the day with the largest differences in temperature effects between lactation days.
Sensitivity against temperature appears to increase over lactation days, but stabilizes earlier for young compared to older sows.

\section{Discussion}
\label{sec:discussion}
\noindent
This work was motivated by the study of heat stress
effects on lactating sows. We used a model framework for scalar-on-function quantile regression for clustered or longitudinal data where dependence within cluster/subject is taken into account by including cluster- or subject-specific intercept parameters. 
Estimation relies on basis expansions, more specifically penalized splines. As an alternative, an eigenfunction basis could be used, with the number of basis functions selected with an AIC criterion (inspired by \cite{kato2012estimation}). This was studied in \cite{phd_thesis} and gave more wiggly estimates of $\beta^\tau(\cdot,\cdot)$, compared to those in Figure~\ref{fig:coef_splines_betast}. 
We prefer the spline expansions over the eigenfunction expansions, in
particular because it avoids the extra step with selection of the
size of the basis.
We adapted existing software, making the methodology more easily applicable for practitioners, see the appendix for implementation details and the supplementary materials fore example code.

Our analysis helped uncover some interesting insights for the sow data application. First, the feed intake quantiles are similar for younger and older sows close to giving birth, but increase faster and to a higher level for older than younger sows, suggesting that sows at their second or later pregnancy acclimatize faster to the environment.  
Second, a high temperature in the stable affects feed intake negatively except for the early days in the lactation period; this is the case for both younger and older sows, and both at the median and at the 0.1 quantile levels. At early lactation days, the temperature effect is not significant, and also similar for both groups of sows.
Third, the estimated temperature effect is generally larger at the 0.1 level compared to the 0.5 level, suggesting that the lower tail of daily feed intake for a sow is more sensible to variation in temperature; however this should be investigated further. 
Fourth, there is an increasing trend of the temperature effect throughout the lactation period at the 0.1 quantile level, and steeper for the older sows.
This is confirmed by model comparisons where models with time-varying temperature effect are preferred over models with constant temperature effects. Fifth, for both groups and at the 0.1 quantile level, the model with functional effect of temperature over the day is preferred over a model which includes the average temperature only. This suggests that the shape, not only the level, of the temperature profile affects the fraction of sows with a low feed intake.

We have focused on models with a single functional covariate, but they could be extended to include more than one functional covariate or a mixture of functional and scalar covariates in a straight-forward way. Moreover, several grouping levels could be included as random effects; this could be relevant in the application because sows were kept together in the stables. 
It remains to study the robustness of estimates in such more complex models. The proposed models have similar flavor as models from \cite{FDboost}, but we were not able to get reliable estimates with the accompanying software.

We adjusted estimates and standard errors with bootstrap methods. The sampling schemes have been used and studied in simpler models \citep{galvao2015bootstrap, battagliola2022}, but further examination would be interesting in the current set-up. Another future research topic
is the development of hypothesis testing procedures for the regression coefficient functions; see \cite{li2022inference}. The recent approaches of \citet{abramowicz2018} and \cite{Pini2023} might be helpful for this purpose. Preliminary ideas involve test statistics computed as integrals over the domain $\mathcal{S}$ of pointwise test statistics and bootstrap computations for evaluation of their null distributions. The main challenge lies in designing appropriate permutation schemes that comply with the dependence structures in the data.

\section{Data availability}
\noindent
The datasets generated during and/or analyzed during the current study are not publicly available due to write proprietary reasons but are available from the corresponding author on reasonable request.
\section{Funding and conflicts of interests
}
\noindent 
The project was partly funded by the Danish Research Council (DFF grant 7014-00221). Moreover, the authors declare they do not have any conflict of interest.
\section*{Appendix}
\label{sec:appendix}

\color{black}

\subsection*{Bootstrap schemes}
\noindent
We detail the resampling schemes mentioned in Section \ref{sec:boot} and used to compute part of the results in Section \ref{sec:application_predQ0}.

\subsubsection*{Block resampling (for assessment of sampling variation of estimator)}
\noindent
Let $c_1^*, \ldots, c_N^*$ be sampled with replacement from the index set of clusters, $\{1,\ldots,N\}$, and define the bootstrap dataset as $\{(Y_{c^*_ij},X_{c^*_ij}(s_h), t_{c^*_ij})\}_{ijh}$. In this way, the within-subject dependence is maintained. For a target, $\theta$, defined from model parameters, we proceed as follows: Draw a bootstrap sample as just described, carry out estimation, and compute the estimated target. Repeat this $B$ times, and denote the estimates $\theta_1,\ldots,\tilde \theta_B$. Finally, compute the standard deviation over the bootstrap estimates as stated in the main text. 
The same method was used by \cite{doi:10.1111/j.1368-423X.2011.00349.x} and
\cite{geraci2014linear}.

\subsubsection*{Combination of standard resampling and wild bootstrap (for assessment of bias)}
\noindent
A bootstrap dataset consists of
$\{(Y_{ij}^*,X_{ij}(s_h), t_{ij})\}_{ijh}$ where
\begin{equation}
\label{eq:Ystar}
    Y_{ij}^* = \hat\alpha^\tau(t_{ij}) + \int_\mathcal{S} \hat\beta^\tau(s,t_{ij}) X_{ij}(s)\, ds + \varepsilon_{ij}^* + u_i^*. 
\end{equation}
The estimates $\hat\alpha^\tau(\cdot)$ and $\hat\beta^\tau(\cdot,\cdot)$ are based on the observed data.
Notice that the values of $X_{ij}(s_h)$ and $t_{ij}$ from the observed data are used unchanged. 
The subject-specific intercepts $u_1^*,\ldots,u_N^*$ are drawn with replacement from the estimates $\hat u_1,\ldots,\hat u_N$ obtained from the observed data, and the error terms $\{\varepsilon_{ij}^*\}_{ij}$ are generated via wild bootstrap. This means that $\varepsilon^*_{ij}=w_{ij}|\varepsilon_{ij}|$, where $\varepsilon_{ij}= Y_{ij} - \hat\alpha^\tau(t_{ij}) -\int_\mathcal{S} \hat\beta^\tau(s,t_{ij}) X_{ij}(s)\, ds -\hat u_i$ are residuals from the model, and   $w_{ij}$s are drawn independently as
\begin{equation*}
w_{ij}=\left\{
\begin{array}{@{}ll@{}}
2(1-\tau), & \text{with probability}\ 1-\tau \\
-2\tau, & \text{with probability}\ \tau
\end{array}\right. 
\end{equation*}
For a target of interest, $\theta$, and estimates $\theta_1,\ldots,\tilde \theta_B$ computed from $B$ bootstrap samples, the bias of $\hat\theta$ is estimated as explained in the main text.
Wild bootstrap was introduced by \citet{wu1986} and \citet{liu1988} for mean regression, and adapted to quantile regression by \citet{10.1093/biomet/asr052}. Results in  \citet{10.1093/biomet/asr052}, \cite{wang2018} and \cite{battagliola2022} indicate that wild bootstrap captures asymmetry and heteroskedasticity better than ordinary resampling of residuals.
\subsection*{Implementation}

\noindent
We used the software environment \texttt{R} \citep{R} for the computations. The FACE method used for smoothing is implemented in the function \texttt{fpca.face}, which is part of package \texttt{refund} \citep{refund}. It can handle functional data observed on a dense or a sparse grid and also allows for missing values. One specifies either the selected PVE (\texttt{pve}) or the number of principal components (\texttt{npc}) of choice. The resulting eigenfunctions, the functional mean, and the predicted/smoothed functions are evaluated and returned at a dense grid. 

In order to fit fQGAM, we rely on the package \texttt{qgam} \citep{qgam_Rpaper}. It includes the \texttt{qgam} function, which is a wrapper of the function \texttt{gam} \citep{wood2017generalized, gamm4}. The call to \texttt{qgam} has the following structure:
\begin{verbatim}
   qgam(y ~ formula, qu=tau, data=data)
\end{verbatim}
where \texttt{y} is the response and \texttt{formula} specifies any ordinary covariates, smooth effects, and random effects to include in the model. The quantile level of interest $\tau$ is passed to \texttt{qu}, and the entry \texttt{data} specifies the data frame of interest. The \texttt{formula} for \texttt{qgam} works with the same syntax as for \texttt{gam}. Hence, smooth terms are included with \texttt{s()} in the univariate case and with \texttt{te()} in the multivariate case. In both cases the user can choose the type and the number of basis functions by passing the arguments \texttt{bs} and \texttt{k}, respectively. Importantly for this paper, the functional covariate is included in the smooth term with the option \texttt{by = Xhat}, where 
\texttt{Xhat} is the matrix of (possibly) pre-smoothed functional covariate values, 
along with the grids of times points of observation, in form of a matrix. The subject-specific intercept is included as a smooth term with \texttt{bs='re'}.

The output from \texttt{qgam} is a \texttt{gamObject},
which stores several quantities related to the model and the estimation process, 
such as
twice the log-likelihood (\texttt{logLik}) and the estimated effective degrees of freedom (\texttt{edf2}), which are used for computation of AIC values in our application.
Moreover, \texttt{Vp}, the variance-covariance matrix of all estimated coefficients, is available, and can be used to compute model-based standard errors and confidence bands for functions of the parameters, such as the targets mentioned in Section~\ref{sec:boot}.  
Estimated quantiles for new covariate functions can be computed with the function \texttt{predict.gam}, and the function allows to exclude one or more terms from the model, such as the subject-specific intercepts, in the prediction.

Finally, we have implemented the bootstrap schemes described in Section \ref{sec:boot}, namely block bootstrap and wild bootstrap, and the \texttt{R} functions are available as part of Supplementary Material in Section S3. Moreover, data preparation, implementation of fQGAM and bootstrap-based inference results are provided in an example using a dataset which is available in the \texttt{refund} package \citep{refund}.

\bibliographystyle{apalike}

\bibliography{bibliography.bib}

\end{document}